\documentstyle[11pt]{article}
\input epsf
\input psfig

\begin{document}


\thispagestyle{empty}
\normalsize

%
%
%

\title{
\begin{flushright}
{\large {\bf UTEXAS-HEP-98-5} \\
		{\bf DOE-ER40757-111} \\
		{\bf UTKL-134} \\}
\end{flushright}
\vskip0.50in
A  Technique of Direct Tension Measurement \\
of a Strung Fine Wire \\
		\vskip0.25in
		\vskip0.25in
}
\author{K. Lang, J. Ting, V. Vassilakopoulos 
\\ 
\\
\it Department of Physics \\
\it The University of Texas at Austin \\
\it Austin, Texas 78712-1081}


\maketitle

\begin{abstract}
\noindent

We present a new technique of direct measurement of 
wire tensions in wire chambers. 
A specially designed circuit
plucks the wire using the Lorentz force
and measures the frequency of damped transverse oscillations of the wire.
The technique avoids the usual time-consuming necessity of tuning 
circuit parameter to a resonance. 
It allows a fast and convenient determination 
of tensions and is straightforward to implement.
\end{abstract}

\thispagestyle{empty}
\newpage
\section{Introduction}

Many techniques of measuring wire tensions in wire chambers 
are extensively covered in the 
literature~\cite{Borghesi}$-$\cite{AtlasTRT}.
Here, we are presenting a new scheme of
performing this task inspired by the principle of generating
sound by vibrating strings in an electric guitar.  
In our novel circuit the plucking is
accomplished by forcing a voltage pulse
through a wire placed in a static magnetic field.
A plucked wire strung in a chamber undergoes damped transverse
oscillations with frequencies given by the formula:
\begin{equation}
\rm f = {n \over 2 L}\sqrt{ ~{T \over \rho} } \label{Equation_1}
\end{equation}
where f is the frequency of oscillations, 
L is the wire length, $\rho$ is the linear density
of the wire, T is the tension force stretching the wire,
and n is the order of the harmonic. 
Frequency of these transverse oscillations
can be directly measured
by analyzing a waveform of the induced Faraday current.

The main difference between the method presented here and all 
other techniques known to us
is that our circuit offers {\it direct} measurement of vibrational
frequency (up to a calibration factor)  
without a usual {\it necessity of tuning} the instrument 
to the resonant frequency of a wire. 
Once a wire is placed in a magnetic field and is connected to the circuit,
the period of oscillations - thus the tension - is determined
automatically. 
This new technique allows the tension measurements to 
be performed significantly faster than using other methods.
We have conducted many tests validating this technique
using wires with lengths ranging from 5~cm to 100~cm.
Our measurements were done in various conditions 
and show an excellent agreement between 
the expected and measured oscillation frequencies.
In the following sections,
we first briefly review the basic principles of the most commonly used
setups for measuring wire tensions, 
and then describe in detail our technique and 
performed tests.

\section{ A Brief General Overview}

The development of various wire chambers since early 1970's and
the task of manufacturing reliable long-lived high-quality instruments
motivated many different approaches to measuring tensions of wires.
Over the years several types of methods have been developed
differing in employed physics principles. They range from 
the most obvious and commonly used electromagnetic and 
electrostatic techniques  to optical measurements.
In this section we give a brief overview of various methods.

A straightforward but somewhat cumbersome way of
determining the tensions may be accomplished
by measuring the sagitta of a horizontally strung
wire. A known weight hung in the center between the two wire-end
supports displaces the wire by a distance which depends on the force 
stretching the wire. This distance may be
measured by an electro-mechanical position transducer
such as a small differential transformer~\cite{Borghesi, Cavalli}. 

Usually, more indirect methods are employed in which the wire is first
forced to vibrate and then these vibrations are used to determine
the tension. The wire ``excitation'' can be produced mechanically
by gas jets~\cite{Stephenson}, sound waves, 
or most commonly by electromagnetic or electrostatic forces.
The most popular electromagnetic methods use the Lorentz force
to generate wire oscillations.
Short bursts of sinusoidal or triangular current wave are driven through
the wire placed in a magnetic field. By {\it adjusting} the frequency of
the bursts, the resonance of the wire can be reached.
At that point the amplitude of the oscillations, hence
the induced emf, reaches the maximum, and the driving and induced
waveforms have the same phase
~\cite{Asano, Bhadra, Calvetti, Coupland, Hoshi, Regan}.
The determination of the resonant frequency
is usually accomplished by displaying
the driving and the emf waveforms on an X-Y scope.
For the lowest and typically the strongest
resonance the resulting Lissajous figure becomes a straight line.
If induced signals are large, it is also feasible to detect
higher harmonics
instead of the fundamental emf frequency of the resonance~\cite{Bhadra}.

In the electrostatic methods the wire is forced to oscillate
through the Coulomb force. Applying a high voltage sine wave of
variable frequency to an electrode, usually an aluminum plate
placed alongside the wire,
generates mechanical vibrations
~\cite{Borghesi, Cavalli, Burns, Carlsmith, Durkin, Jones}.
To increase the coupling forces, some methods use two electrodes
exactly out of phase with the wire in between~\cite{Carlsmith}.
Other techniques, instead of plate electrodes, employ
parallel neighboring wires as drivers~\cite{Jones}.
As before, at the resonance the wire reaches maximum amplitude
and exhibits the same phase as the driving waveform.
These techniques do not employ magnetic field, thus there is no emf and
the determination of the resonance can be done visually,
if the wire is visible~\cite{Borghesi, Cavalli},
or indirectly by observing the waveform generated by the
change of capacitance between the wire and 
the electrode~\cite{Durkin, Jones, AtlasTRT}.

Generally, the electrostatically 
coupled signal is small so that sensitive balanced bridges
need to be used for measurements~\cite{Cavalli, Shenhav}.  
In a balanced impedance bridge, the signal is fed back by 
an amplifier in a circuit analogous to that of a crystal 
oscillator~\cite{D'Antone}. In this way, the 
non-linear impedance of the wire guarantees that 
the circuit oscillates at the natural mechanical 
resonance frequency of the wire. However, as it 
is in a typical bridge oscillator, the values of 
the components have to be tuned very close to 
the resonance or the circuit will not oscillate.

It should be pointed out that in all the above methods
the circuits need to be {\it tuned} to the resonant
frequency of the wire in order to measure the tension.
In most cases this is time consuming and requires
careful adjustments of the driving pulses and
sensing circuitry. The technique which we are
proposing here evades these deficiencies. Our circuit
{\it directly determines} the resonant frequency by analyzing
the characteristic damped oscillation waveform.
The measurement does not require any tuning, is fast and robust.

\section{Circuit Concept}

\begin{figure}
\epsfxsize=4.3in
\centerline{ \epsfbox{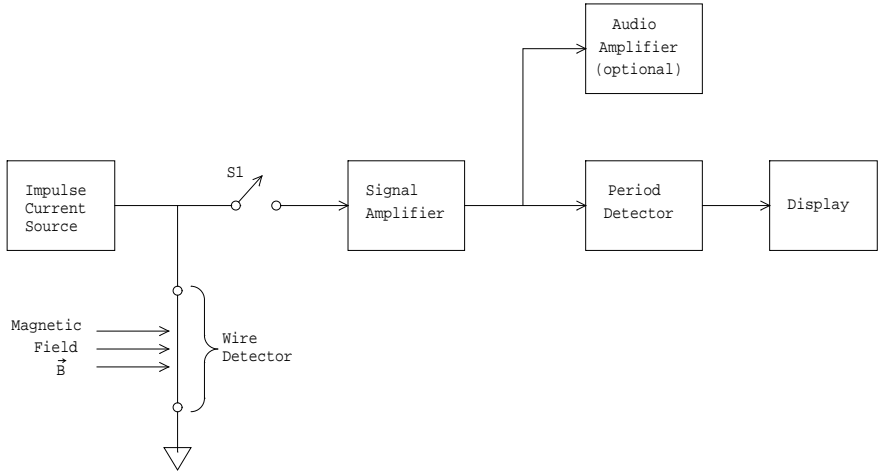}  }
\caption{\footnotesize 
A conceptual diagram of the circuit for
direct measurements of wire tensions. 
\label{Conceptual_Figure} }
\end{figure}

The circuit which we designed is conceptually very simple,
as illustrated in Figure~\ref{Conceptual_Figure}.
It consists of an impulse current source, followed by a signal amplifier 
and a frequency detector. An optional acoustic amplifier was added 
to further facilitate the measurements. 
The principles behind the
circuit are the Lorentz force to excite the wire mechanically
and the Faraday law to induce a signal on the wire 
vibrating in a magnetic field.
We have primarily tested it using a straw tube wire chamber,
but it could be used for most wire chambers. 
We found that a conductive straw,
acting as a Faraday cage, provides a better noise immunity.
Our reference description below refers
to a wire strung in a 5mm-diameter copperized 
Mylar straw tube~\cite{Graessle}.

To measure the oscillation frequency
the far-end of the wire is shorted to the ground potential. 
Both the current excitation and the induced 
signal readout are done from the near-end.  
A permanent magnet is placed
near the mid-point of the wire.  
At the excitation stage a current pulse (of magnitude of the order 
of tens of milliamperes) is forced through the wire.
It does not flow through the detection circuit
because the analog switch S1 is open. This current generates a Lorentz
force which is both orthogonal to the wire and to the magnetic field, 
thereby mechanically displacing the wire.
After the current is turned off,
the wire vibrates with damped harmonic oscillations
and ultimately returns to its rest position.
The oscillations of the wire in the magnetic field
generate a Faraday current with classical exponential decay envelope
which flows through the now-closed switch S1,
and into the low impedance measuring part of the circuit. It is first
amplified, then filtered, and then the period of the oscillations is
determined and displayed on an LCD panel.

\section{Initial Circuit Implementation}

The translation of this simple idea into a practical device 
required following
all the basic low-noise circuit practices including
a solid single point ground and careful shielding.
In the initial trials we used a square voltage excitation pulse.
The frequency of oscillations behaved
erratically as a function of the tension of the wire.
The source of this problem was found to be the interference
of two out-of-phase oscillations induced on the wire.
One was caused by the leading edge of the excitation pulse
and another by its falling edge. The resulting oscillation
was indirectly modulated by the width of the excitation pulse.
The solution was found by changing the shape of
the excitation pulse - a square was replaced by
a saw-tooth waveform current.
Using a smoother leading edge
the initial displacement of the wire is slower, while
the releasing is sudden - mimicking plucking of a guitar string.  

Another problem found was that for wire oscillations in the frequency
range of the order of
60 Hz the interference with the AC power led to somewhat
erratic behavior of the circuit. This problem was solved by
placing the power supplies and the transformers beneath
the circuit box, and by adding a ferrous magnetic shield 
plate in between.  

\section{Detailed Circuit Description}

\begin{figure}
\epsfysize=5.8in
\centerline{ \epsfbox[90 0 450 600] {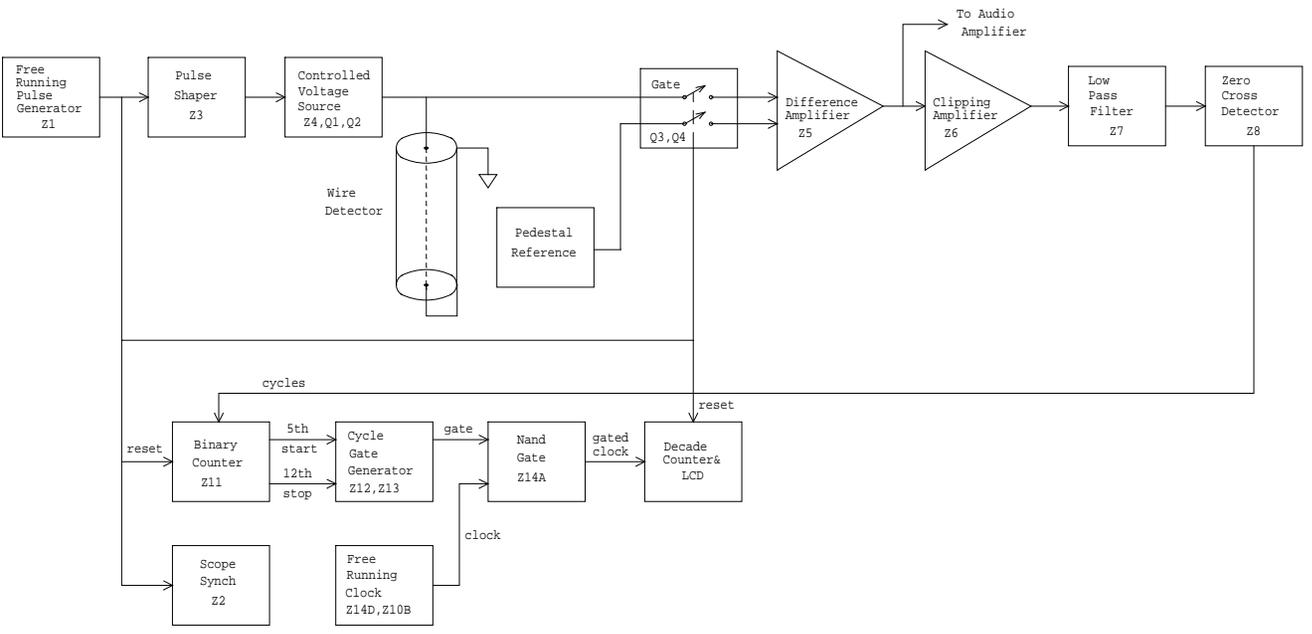} }
\vskip0.3in
\caption{  \footnotesize A block diagram of the designed circuit to
directly measure wire tensions. \label{BlockDiagram} }
\end{figure}

\begin{figure}
\epsfysize=5.8in
\centerline{ \epsfbox[90 0 450 600] { 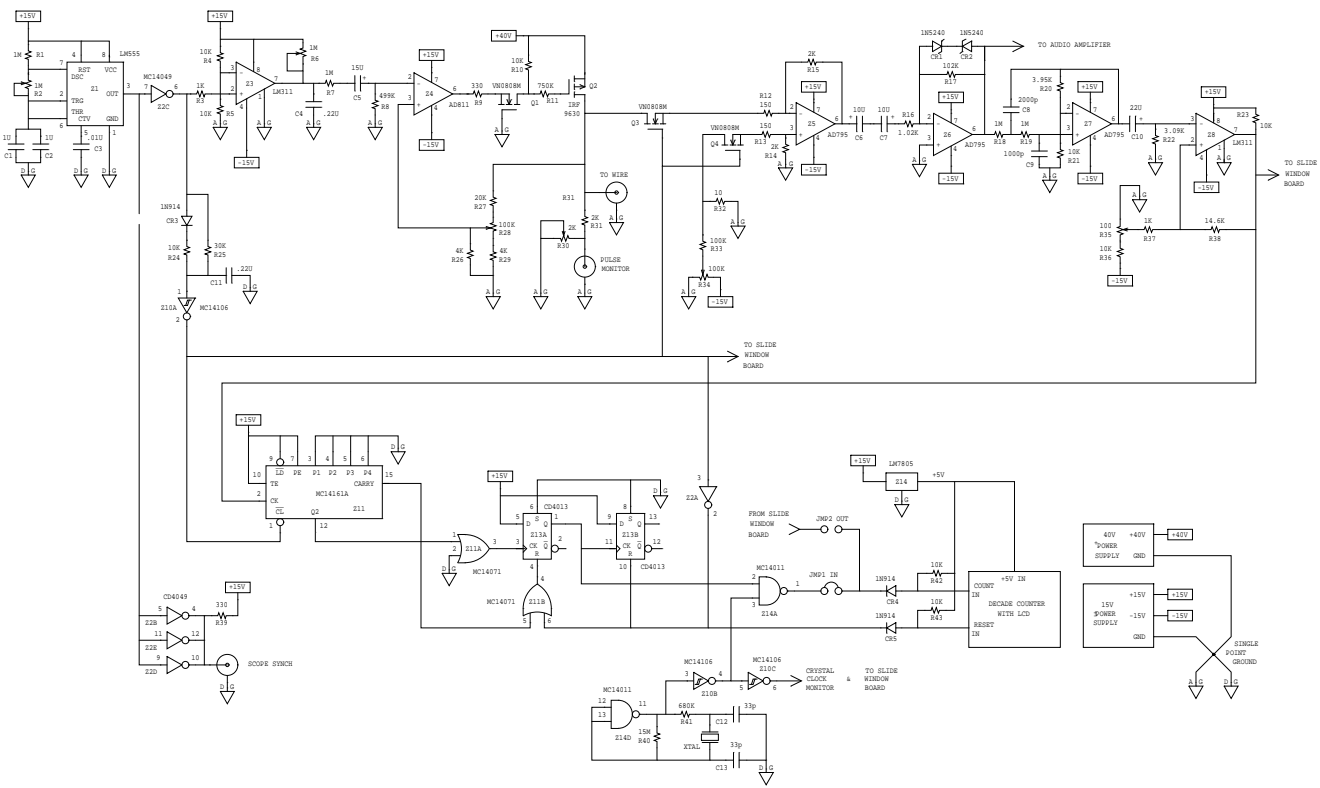}  }
\vskip0.3in
\caption{  \footnotesize A detailed schematics of the designed circuit to
directly measure wire tensions. \label{DetailSchematic} }
\end{figure}

The circuit, shown in block diagrams in Figure~\ref{BlockDiagram},
can be divided into several main functional blocks:
Free Running Pulse Generator,
Pulse Shaper,
Controlled Voltage Source,
Analog Switch,
Difference Amplifier,
Clipping Amplifier,
Low Pass Filter,
Zero Crossing Detector,
Binary Counter,
Cycle Gate Generator,
NAND gate,
Decade Counter,
LCD,
Free Running  Crystal Clock,
and Scope Synch.
Below we discuss the functions of each part and refer to the detailed
circuit schematics shown in Figure~\ref{DetailSchematic}.

\begin{figure}
\epsfxsize=5in
\epsfysize=1.75in
\centerline{ \epsfbox{ 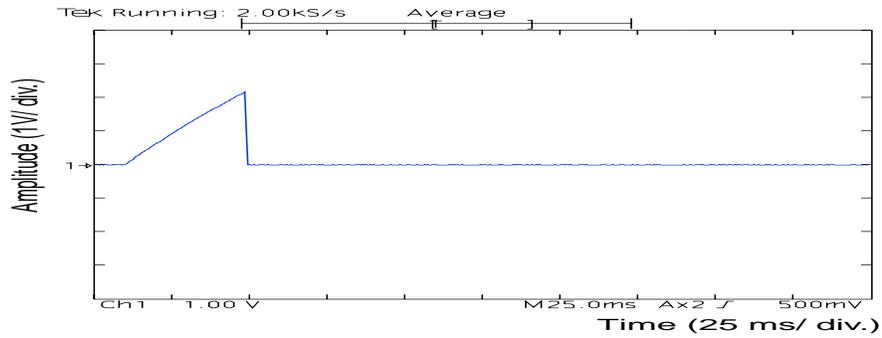}  }
\caption{ \footnotesize 
	An oscilloscope trace showing the shape of
	the wire plucking excitation pulse. 
\label{ExcitationPulse} }
\end{figure}

The Free Running Pulse Generator Z1 is a 555 timer chip wired as a free
running astable multi-vibrator oscillating at about 1~Hz.  The Pulse Shaper 
changes the fast rising edge of the rectangular pulse into a slowly rising
exponential with the R6C4 time constant.  
The trailing edge remains fast, as shown in Figure~\ref{ExcitationPulse}.
The Controlled Voltage Source changes the reference voltage pulse
into a power voltage source.
The peak wire voltage can be adjusted up to 10:1 ratio
by varying R28. This adjustment improves the signal-to-noise
ratio. We typically used peak voltage of 18~V, which corresponds to
62~mA current for the 20~$\mu$m diameter gold-plated 
tungsten wire which we used~\cite{Luma}.
The Analog Switch consists of a pair of
FET switches Q3 and Q4.  Q3 controls the signal path.  Q4  switches a DC
reference pedestal.  When both are closed at the same time, the switching
transients are canceled out by the Difference Amplifier Z5.  
It is followed by an AC amplifier Z6, which
slightly differentiates the signal in order to balance 
it around ground, and also clips large signals to prevent the
saturation of the op amp.  The clipping is provided  by a pair of back
to back 10 V zener diodes CR2 and CR3.

\begin{figure}
\epsfxsize=5in
\epsfysize=1.75in
\centerline{ \epsfbox{ 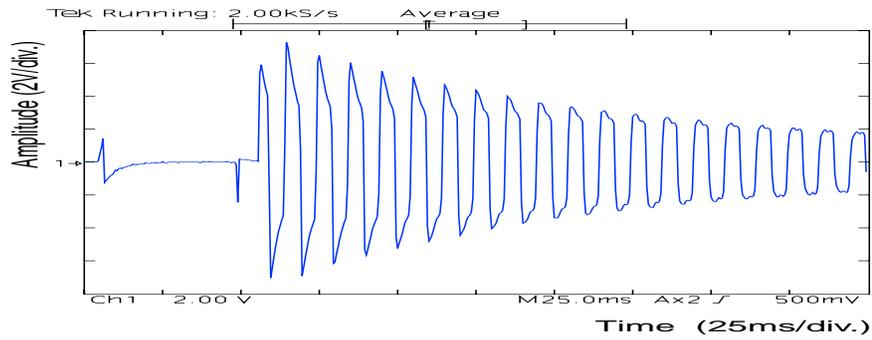}  }
\caption{  \footnotesize 
	The oscilloscope trace of the induced signal from the oscillating
	wire. The signal is not exactly damped sinusoidal due to short range
    of the magnetic field.
\label{RawWave} }
\end{figure}

The Low Pass Filter lets through 
only the  fundamental frequency (below 200 Hz) 
and attenuates all higher harmonics. Depending on the
application this filter may need to be adjusted or removed.
The Zero Crossing Detector transforms the exponential 
decaying sine wave, as shown in Figure~\ref{RawWave},
into a series of logic pulses. 
The period of the 
oscillations is measured by a Free Running Crystal Clock. 
The time window selected for the period measurement
is produced by the Cycle Gate Generator and begins 
at the 5th and ends with the 12th cycle.  Thus, 8 complete cycles are gated 
through and measured by the Crystal Clock.  The CMOS Decade Counter 
with its 6-digit LCD is a modified ``Red Lion'' counter 
where only the 4 most significant digits are used.  This arrangement 
yields good low noise results. A typical reading  of about 
2011 counts has a $\pm$~2 counts uncertainty.
For further convenience we have also added
an audio amplifier 
which translates the measured frequency to a speaker sound.

The Crystal Clock runs continuously at a measured
frequency of 32,770 Hz 
(2~Hz higher than the nominal 32,768~Hz).
The frequency of the wire oscillations, f,
(to be used in formula~\ref{Equation_1}) is determined by
a simple expression: 
\begin{equation}
\rm f = {32,770 \over  {^{N_{count}} / _{8} } } ~~[Hz] \label{Equation_2}
\end{equation}
where $\rm N_{count}$ is the number of clock counts within the
measuring time window and it is displayed on the LCD panel, 
8 stands for the number of cycles within the measuring time window, 
and 32,770 is the effective frequency of the clock.

The ``Red Lion'' counter and the LCD require a 5~V power supply.  
All other logic units use 15~V CMOS primarily 
for good noise immunity and the capability of direct interface 
with power MOSFETs.

\begin{figure}
\epsfxsize=5in
\epsfysize=1.75in
\centerline{ \epsfbox{ 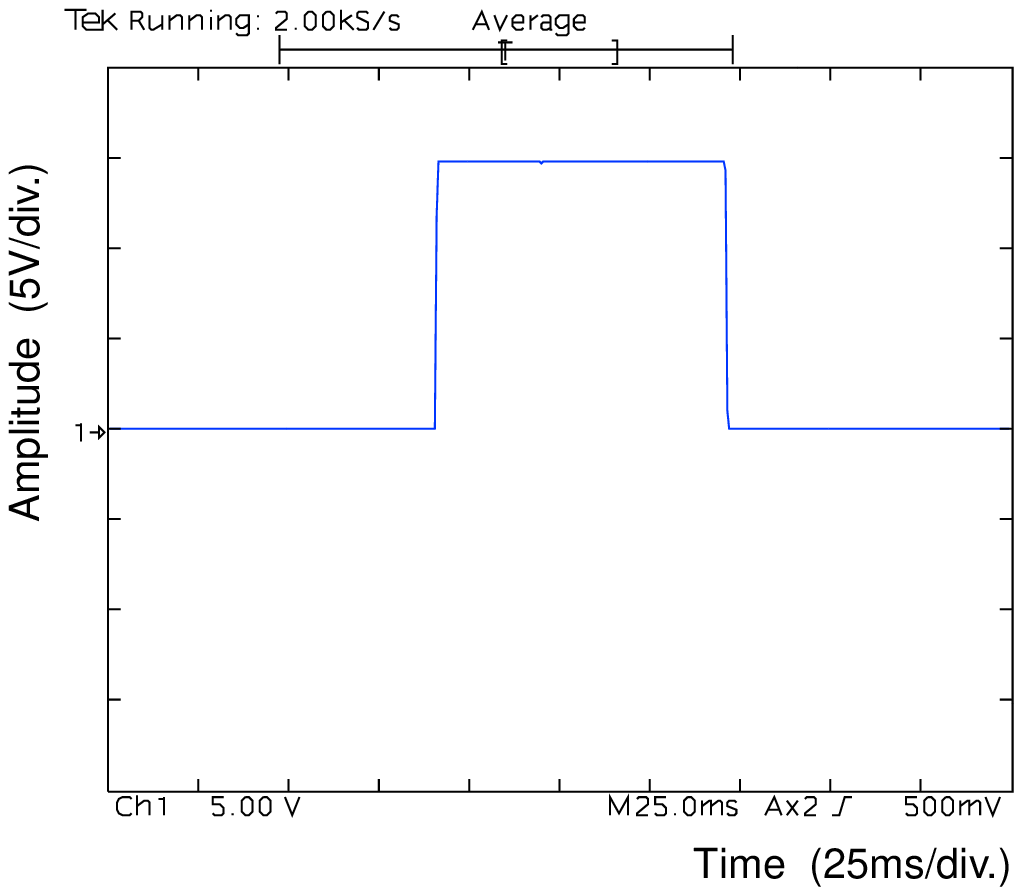}  }
\caption{  \footnotesize
	The oscilloscope trace of the signal from the Cycle Gate Generator.
	The gate is spanning exactly 8 periods (5th through the 12th)
	is measured to detect the number of zero-crossings. 
\label{Gate} }
\end{figure}

\section{Test Results}

To determine the functioning of the circuit many
measurements were conducted with a variety of circuit parameters
and wire conditions. In most tests we used a gold-plated tungsten
wire with a 20~$\mu$m diameter and a measured linear density of
$\rho =$ 61 $\mu$g/cm~\cite{Luma}.
For our tests the wire was secured at different tensions 
by soldering it to two vertically spaced fixed points.
The wire was first soldered at the top fixed point,
then a weight was attached to it and after the wire
stabilized it was soldered to the bottom fixed point which
was also electrically in contact with a BNC signal connector.
Every time the weight changed the solder was made to re-flow.
The conversion of the measured periods into frequencies was done 
``offline'', but if desired it could be hard-wired into 
a computer or a microprocessor to convert the direct readings into 
frequencies or tension. 

First, we experimentally verified that
the strength of the magnetic field changed only
the amplitude of the signal and not
its frequency, even when the signals were very large and
thus heavily clipped.
The position of the magnet along the wire did change the harmonic 
content of the signal, but 
not its fundamental frequency, and 
therefore the results were essentially the same.  When the magnet 
is placed at mid-way along the wire the fundamental frequency dominates,
the signal-to-noise ratio reaches the highest value,
therefore this magnet position was kept 
throughout the measurements.  It should be also pointed out that
the induced emf is not exactly sinusoidal but rather has a somewhat 
square-wave shape as shown in Figures~\ref{RawWave}.  
The cause of this effect is the relatively short range of the
applied magnetic field as compared to the length of the wire.
The triangular wave shape is the result of the differentiation of 
the square wave.
When a single magnet was replaced by 
a pair of separated magnets along the wire, 
the signal became much more sine-wave like.  
Figures~\ref{ExcitationPulse}, \ref{RawWave}, and \ref {Gate}
show typical synchronized oscilloscope traces of the excitation pulse, 
the induced damped emf, and the standard time window for zero-crossing
measurement.

\begin{figure}
\epsfxsize=5.0in
\centerline{ \epsfbox{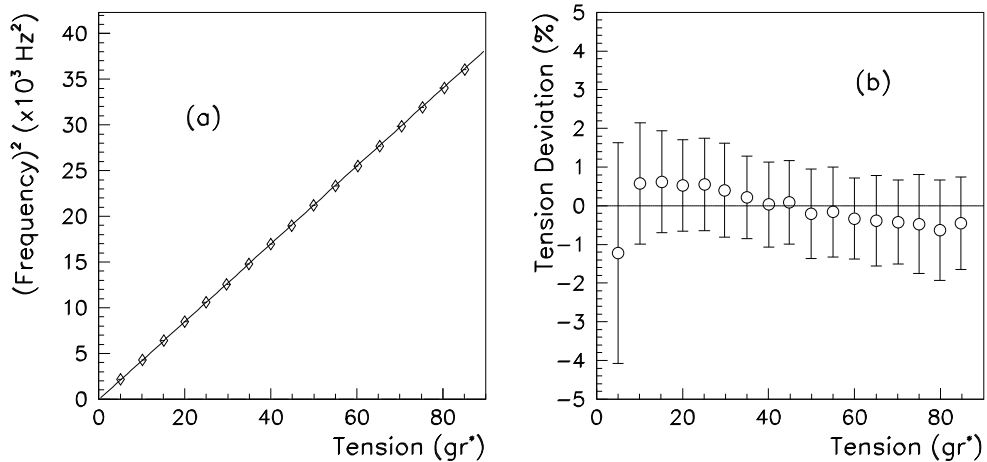} }
\caption{  \footnotesize (a) Measured frequency squared
of a test wire as a function of tension. The points with error bars
were obtained using a 1~m long 20~$\mu$m gold-plated tungsten wire.
The straight line is the predicted fundamental frequency. The excellent
agreement is further illustrated in (b) where the same results
are plotted as percentage of deviation from the expected value.}
\label{Calibration} 
\end{figure}

The tension of the wire was measured for a wide range of tensions
and up to the breaking force of
the 20~$\mu$m diameter gold-plated tungsten wire of about 1.2~N. 
The results are shown in Figures~\ref{Calibration}(a) and (b).
The error bars reflect the uncertainty in the wire density, the wire
length, and the standard deviation of 10 readings on the instrument.
An excellent agreement between the predicted
and measured frequencies has been achieved with the circuit.

\begin{figure}
\epsfxsize=4.3in
\centerline{ \epsfbox{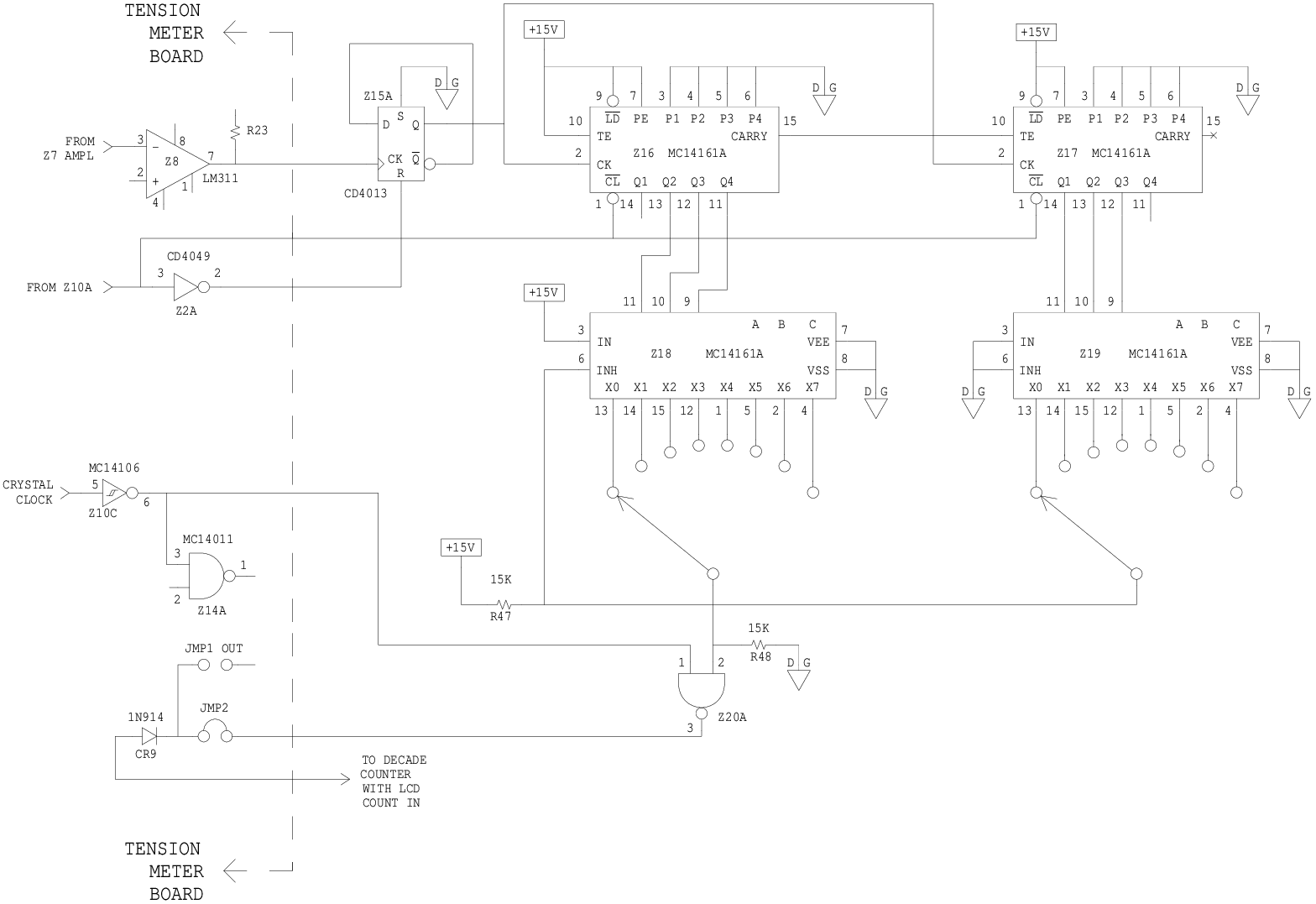}  }
\caption{  \footnotesize  
	A circuit for the sliding time window test. 
\label{SlidingWindowCircuit} }
\end{figure}

\begin{figure}
\epsfxsize=4.0in
\centerline{ \epsfbox{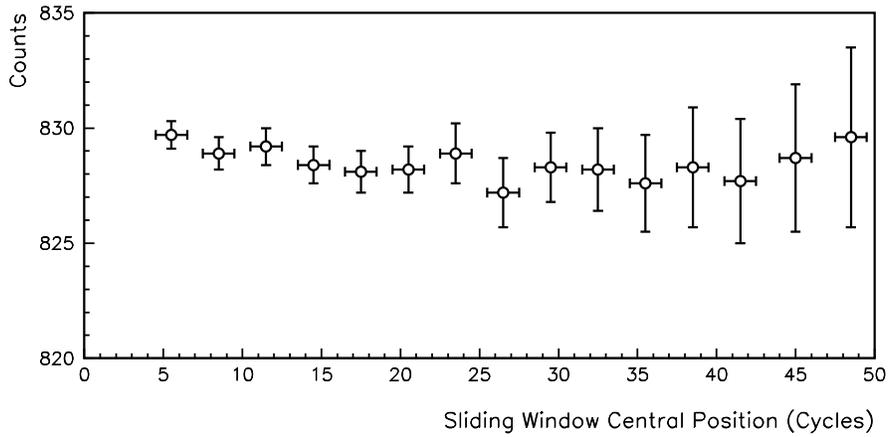}  }
\caption{  \footnotesize   Measured frequency (counts) of a test wire
as a function of the time window used.  The uncertainty increases
if later zero-crossings are used. 
\label{SlidingWindowResults} }
\end{figure}

We also performed a test to check whether the determination of
a period of oscillations depends on the position of the time window used.
A special circuit generating a sliding measurement time window
was built and is shown in Figure~\ref{SlidingWindowCircuit}.
This additional circuit allowed the measurement of the number of
zero-crossings between the 4th and the 7th cycle, or between the
7th and 10th, or 10th and 13th,  etc., at every 4 cycle intervals.
Unfortunately the circuit was too big to fit inside the
original circuit box.  Its use decreased the signal to noise
ratio of the system therefore it was removed after this test.
The averages and standard deviations were taken from
50 consecutive readings
for each window position.
The measurements indicate that the readings are
fairly independent of the time window.
The standard deviation of the measured frequency increases slightly
with the time interval between the beginning of the
oscillations and the time position of the measuring window.
This is expected because the amplitude of the oscillations
decreases with time, and consequently its signal-to-noise
ratio worsened.
The results of these tests are shown in
Figure~\ref{SlidingWindowResults}.

In our test we liked the addition of an audio amplifier, 
as implemented in reference~\cite{Coupland}, which
significantly facilitated  
the tediousness of the measurements.  
After a bit of practice, 
one can detect the abnormalities of the wire like shorts, breakage,
or very low tensions 
just by the sound, and without reading the frequency on the LCD monitor.  
For a large set of wires the tension measurements can be done fast 
by first grounding all ends on one side, and
then by connecting successively the other end of each wire to the circuit.
In our tests it typically took only few seconds per channel
to determine the wire tensions.

\section{Summary}

We have designed and tested a new circuit to measure tensions of
wires in wire chambers. Our simple technique of plucking the wire
electromagnetically and analyzing the frequency of the induced signal
is fast, robust and easy to implement. It offers a direct method
of determining the tensions without a time-consuming necessity to tune
the circuit to a resonance. Our tests showed better than 1\% agreement 
between the applied and measured tension values.



\begin{thebibliography}{99}
\bibitem{Luma}
	We have used a 20~$\mu$m diameter
	gold plated tungsten wire \#821 manufactured by
	Luma Metall International, Kalmar, Sweden.
\end{thebibliography}
\end{document}